\documentclass[aip,jcp,preprint]{revtex4-1}

\usepackage{graphicx}
\usepackage{longtable}
\usepackage{array}
\usepackage{dcolumn}
\usepackage{amsmath}
\usepackage{amsfonts}
\usepackage{amssymb}
\usepackage{soul}

\usepackage{color}

\begin{document}

\title{Interplay of spin-dependent delocalization and magnetic anisotropy in the ground and excited states of [Gd$_2$@C$_{78}$]$^{-}$ and [Gd$_2$@C$_{80}$]$^{-}$}
\author{Akseli Mansikkam\"aki}
\email{akseli.mansikkamaki@jyu.fi}
\affiliation{Theory of Nanomaterials Group, Chemistry Department, Katholieke Universiteit Leuven, Celestijnenlaan 200F, 3001 Leuven, Belgium.}
\affiliation{Department of Chemistry, Nanoscience Center, University of Jyv\"askyl\"a, P.\,O. Box 35, Jyv\"askyl\"a, FI-40014, Finland.}

\author{Alexey A. Popov}
\affiliation{Leibniz Institute for Solid State and Materials Research Dresden, Helmholtzstrasse 20, Dresden 01069, Germany.}

\author{Qingming Deng}
\affiliation{Leibniz Institute for Solid State and Materials Research Dresden, Helmholtzstrasse 20, Dresden 01069, Germany.}
\affiliation{School of Physics and Electronic Electrical Engineering, Huaiyin Normal University, Huai'an, 223001, China.}

\author{Naoya Iwahara}
\affiliation{Theory of Nanomaterials Group, Chemistry Department, Katholieke Universiteit Leuven, Celestijnenlaan 200F, 3001 Leuven, Belgium.}

\author{Liviu F. Chibotaru}
\email{liviu.chibotaru@chem.kuleuven.be}
\affiliation{Theory of Nanomaterials Group, Chemistry Department, Katholieke Universiteit Leuven, Celestijnenlaan 200F, 3001 Leuven, Belgium.}

\begin{abstract}
  The magnetic properties and electronic structure of the ground and excited states of two recently characterized endohedral metallo-fullerenes, [Gd$_2$@C$_{78}$]$^{-}$ (\textbf{1}) and [Gd$_2$@C$_{80}$]$^{-}$ (\textbf{2}), have been studied by theoretical methods. The systems can be considered as [Gd$_2$]$^{5+}$ dimers encapsulated in a fullerene cage with the fifteen unpaired electrons ferromagnetically coupled into an $S=15/2$ high-spin configuration in the ground state. The microscopic mechanisms governing the Gd--Gd interactions leading to the ferromagnetic ground state are examined by a combination of density functional and \emph{ab initio} calculations and the full energy spectrum of the ground and lowest excited states is constructed by means of \emph{ab initio} model Hamiltonians. The ground state is characterized by strong electron delocalization bordering on a $\sigma$ type one-electron covalent bond and minor zero-field splitting (ZFS) which is successfully described as a second order spin-orbit coupling effect. We have shown that the observed ferromagnetic interaction originates from Hund's rule coupling and not from the conventional double exchange mechanism. The calculated ZFS parameters of \textbf{1} and \textbf{2} in their optimized geometries are in qualitative agreement with experimental EPR results. The higher excited states display less electron delocalization but at the same time they possess unquenched first-order angular momentum. This leads to strong spin-orbit coupling and highly anisotropic energy spectrum. The analysis of the excited states presented here constitutes the first detailed study of the effects of spin-dependent delocalization in the presence of first order orbital angular momentum and the obtained results can be applied to other mixed valence lanthanide systems.
\end{abstract}

\maketitle

\section{Introduction}

Endohedral metallo-fullerenes (EMFs) can stabilize novel chemical specie such as small lanthanide clusters by the steric protection and the large electron affinities of the fullerene cages. Such small metal clusters can then exhibit unusual magnetic properties.\cite{chen_2015a,popov_2013a,lu_2012a} For example, LnSc$_2$N@C$_{80}$ and Ln$_2$ScN@C$_{80}$systems (Ln = lanthanide) show highly pronounced single-molecule magnet (SMM) behaviour\cite{greber_2012a,popov_2015a,popov_2017b} due to highly axial crystal field around the Ln ions, and Dy$_3$N@C$_{80}$ is a geometrically frustrated magnetic molecule with unquenched orbital angular momentum at the Ln sites\cite{chibotaru_2013a}.

Over the past few years, the magnetic properties of a class of EMFs where [Ln$_2$]$^{5+}$ dimers are encapsulated in various fullerene \cite{kato_2016a,lu_2016a} or azafullerene (C$_{79}$N)\cite{dorn_2008a, dorn_2011a} cages have attracted significant attention as some of them exhibit short Ln--Ln distances bordering on a covalent Ln--Ln bond\cite{popov_2012a} and strong ferromagnetic coupling in their ground spin states. Density functional theory (DFT) calculations on Gd$_2$@C$_{79}$N have shown that the two Gd ions have 4f$^7$ electron configurations and the one ``extra'' electron occupies a $\sigma$ type orbital with significant contributions from 5d, 6s and 6p orbitals.\cite{rajaraman_2015a, daul_2015a} The DFT calculations also show extensive delocalization of the $\sigma$ electron over the two Gd ions and predict strong ferromagnetic exchange interaction leading to an overall $S=15/2$ spin state. At a glance, the ferromagnetic exchange interaction seems to be explained by a double exchange mechanism\cite{zener_1951a,anderson_1955a,de-gennes_1960a} where a 5d electron is resonating between the two Gd ions and is locally coupled to the spins of the Gd 4f electrons. An adequate description, however, is not trivial \emph{a priori} because the electron delocalization coexists with significant spin-orbit coupling (SOC) in the 5d orbitals of the Gd ions. To our knowledge, double exchange interaction in the presence of strong magnetic anisotropy on the magnetic sites has not been addressed so far.

In the present work, we have conducted a theoretical study on the ground and low-lying excited states of two recently characterized ferromagnetic EMFs, [Gd$_2$@C$_{78}$]$^{-}$ (\textbf{1}) and [Gd$_2$@C$_{80}$]$^{-}$ (\textbf{2})\cite{kato_2016a}, the latter of which is isoelectronic to Gd$_2$@C$_{79}$N. The low-temperature EPR spectra of these systems shows that they both have ferromagnetically coupled high-spin $S=15/2$ ground states which show some anisotropy.\cite{kato_2016a} We will analyze the exchange interactions leading to the ferromagnetic ground state and will explain the ground state anisotropy as a second order effect. We will then proceed to discuss the spectrum of the excited states where an electron is promoted to a $\pi$ or $\delta$ symmetric orbital. Due to the strong Gd--Gd interactions imposed by the short interionic distances and large SOC constant of the Gd ions, the spectrum of the excited states is shaped both by strong exchange interactions due to electron delocalization and by strong spin-orbit coupling. The work presented here will also form a solid foundation for the study of more complicated mixed valence lanthanide systems such as Dy analogues of the systems studied here.

\section{Computational details}

\subsection{DFT calculations}

Geometries of \textbf{1} and \textbf{2} were optimized at DFT level using the PBE0 hybrid exchange correlation (XC) functional.\cite{perdew_1996a,perdew_1996b,scuseria_1999a,barone_1999a} Ahlrichs' TZVP \{62111/411/1\} basis set\cite{ahlrichs_1992a} was used for the carbon atoms and ECP53MWB effective core potential of Dolg et al.\cite{preuss_1989a} with \{311111/31111/21111/111/11\} valence part was used for the  Gd ions.\cite{yang_2005a}

All further DFT calculations were carried out on the optimized geometries using all-electron basis sets. Scalar relativistic effects were treated with the zeroth order regular approximation (ZORA)\cite{baerends_1993a,baerends_1994a,baerends_1996a} as implemented in \textsc{Orca}\cite{van-wullen_1998a}. SARC-ZORA-TZVPP\\ \{611111111111111111/511111111111/411111111/411/111\} basis set was used for the Gd ions\cite{neese_2009a} and ZORA-def2-SVP \{511/31/1\} for the carbon atoms\cite{neese_2008a} along with the corresponding auxiliary basis sets. Energetics of low-lying spin states were probed with broken symmetry (BS) DFT calculations using the PBE0 XC functional. Excitation energies into higher-lying states and the splitting of terms under SOC were calculated using the DFT/ROCIS method.\cite{neese_2013a,neese_2013b} The PBE0 XC functional was used along with the default set of parameters for the scaling of the CI matrix elements in the DFT/ROCIS procedure. The CI matrix was constructed by allowing excitations from the singly occupied orbitals into all virtual orbitals up to $5.0$ Hartree (which includes almost all virtual orbitals) and the 200 lowest roots were solved for. SOC effects were introduced by constructing the SOC Hamiltonian in a basis of the DFT/ROCIS eigenstates using the spin-orbit mean-field (SOMF) operator\cite{neese_2005a,palmieri_2000a,hess_1996a} and then diagonalizing it to yield the spin-orbit coupled states and eigenvalues. Quasi-restricted orbitals were used in the DFT/ROCIS calculations instead of actual restricted open-shell orbitals to avoid convergence issues in the restricted open-shell Kohn--Sham calculation which would be necessary to produce the true restricted open-shell orbitals. DFT-based ZFS parameters were calculated with the method proposed by Neese\cite{neese_2006a,neese_2007b} using the pure PBE GGA functional.\cite{perdew_1996a,perdew_1996b}

All DFT calculations were carried out with the \textsc{Orca} program suite.\cite{neese_2012b} Version 3.0.2 was used for the geometry optimizations, 4.0.0 for DFT/ROCIS calculations, and 3.0.3 for all other DFT calculations.

\subsection{Multireference \emph{ab initio} calculations}

Multireference \emph{ab initio} calculations were first performed on a free Gd(II) ion. The single ion calculations  used the ANO-RCC-VQZP basis set, which corresponds to a [9s8p6d4f3g2h] contraction.\cite{roos_2008a} Scalar relativistic effects were included using the exact two-component (X2C) transformation.\cite{kutzelnigg_2005a,filatov_2005a,reiher_2012b} CASSCF calculations were carried out using two different active spaces. An (8,12) space was used to calculate the splitting between the lowest Hund and non-Hund states arising from a 4f$^7$5d$^1$ configuration and a larger (8,16) active space was used to evaluate the relative energies of states arising from 4f$^8$ configurations as compared to the 4f$^7$5d$^1$ states. The (8,12) active space included the seven 4f orbitals and five 5d orbitals and the (8,16) active space also included the 6s and 6p orbitals. A state-averaged CASSCF calculation was carried out to solve the lowest roots in each active space. Five roots with $S=4$ and another five roots with $S=3$ corresponding to the $^9D$ and $^7D$ terms of the Gd(II) ion were solved in the CASSCF(8,12) calculation and a total of nine $S=4$ and sixteen $S=3$ roots corresponding to the $^9D$, $^9S$, $^9P$, $^7D$, $^7S$, $^7P$ and $^7F$ terms were solved in the CASSCF(8,16) calculation. In the CASSCF(8,16) calculation the orbitals were optimized only for the $S=4$ states and the energies of the $S=3$ states were solved by a single diagonalization of the CI matrix to prevent the orbital optimization procedure from rotating 6d orbitals into the active space due to the double shell effect.. The remaining dynamic electron correlation not accounted for in the CASSCF calculations was included as a perturbation correction to the energies using the extended multistate (XMS) CASPT2 method.\cite{roos_1982a,roos_1990a,roos_1992a,werner_2011b} The XMS version of CASPT2 was chosen to avoid unphysical splitting of spatially degenerate states. The XMS-CASPT2 correction was calculated only for the CASSCF eigenvalues and no correction to the wave functions was computed. Finally, SOC was introduced using the restricted active space state interaction (RASSI) methodology\cite{malmqvist_2002a}. The SOC Hamiltonian was constructed in a basis of spin-free states using the atomic mean field integral (AMFI) operator\cite{christiansen_2000a,hess_1996a} and then diagonalized to yield the final spin-orbit coupled states.

CASSCF calculations were also carried out on the full systems \textbf{1} and \textbf{2} using an ANO-RCC-VTZP basis ([8s7p5d3f2g1h] contraction) for the Gd ions and an ANO-RCC-VDZP basis ([3s2p1d] contraction) for the carbon atoms.\cite{roos_2008a,roos_2004b} A minimal (15,15) active space was used which included all 14 4f orbital combinations and a $\sigma$ bonding orbital. One $S=15/2$, two $S=13/2$, two $S=11/2$, two $S=9/2$, two $S=7/2$, two $S=5/2$, two $S=3/2$ and two $S=1/2$ states were solved corresponding to the states in the lowest Hund and non-Hund exchange state in the $\Sigma$ manifolds.

All multireference calculations were carried out with the \textsc{Molcas} quantum chemistry program.\cite{lindh_2015a} The 8.1 development version was used for the XMS-CASPT2 calculations and the rest of the calculations were carried out using the 8.0 release version. All \textsc{Molcas} calculations utilized Cholesky decomposition using a threshold of $10^{-8}$.

\section{Results \& discussion}

\subsection{Geometry optimization}

\begin{figure}
  \includegraphics[width=7cm]{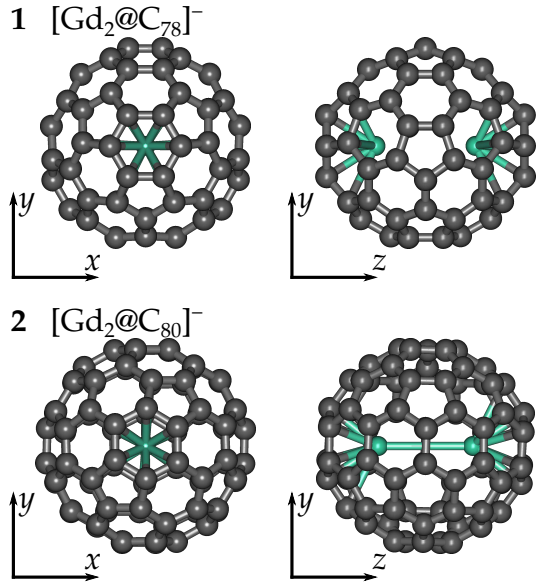}
  \caption{\label{F:geometries}Optimized geometries of \textbf{1} and \textbf{2} (black: C, green: Gd).}
\end{figure}

The geometries of \textbf{1} and \textbf{2} were fully optimized at DFT level.  Details of the optimization procedures are given in Sec. I in the supplementary material and the optimized geometries are presented in Figure \ref{F:geometries}.  \textbf{1} retains the $D_{3h}$ symmetry of an elongated C$_{78}$-$D_{3h}$(5) fullerene cage whereas in \textbf{2} the presence of the Gd ions lowers the symmetry of the C$_{80}$-$I_h$ cage to an approximate $D_{2h}$ symmetry. The symmetry of the latter agrees with that of La$_2$@C$_{80}$ which has been determined by EPR measurements.\cite{kato_2007a} The Gd--Gd distances in \textbf{1} and \textbf{2} are $4.088\,\text{\AA}$ and $3.874\,\text{\AA}$, respectively. The latter distance is similar to the La--La distance of 3.71\,{\AA} observed in the crystal structure of a benzyl adduct of La$_2$@C$_{80}$ which has a cage that is isoelectronic to C${_{80}}^{6-}$.\cite{lu_2016a} These geometries were used in all subsequent calculations.

\subsection{General features of electronic structure}

To get an overall picture of the electronic structures of \textbf{1} and \textbf{2}, DFT calculations were conducted on the high-spin ($S=15/2$) states of both systems. In order to separate features of the electronic structure that are attributable purely to the [Gd$_2$]$^{5+}$ core moiety and to those that result from interaction with the C${_{78/80}}^{6-}$ cage, the calculations were also performed on two simple [Gd$_2$]$^{5+}$ dimers \textbf{1'} and \textbf{2'} which contain only the Gd ions fixed at the same distance as in the respective caged structures \textbf{1} and \textbf{2}.

\begin{figure}
  \includegraphics[width=7cm]{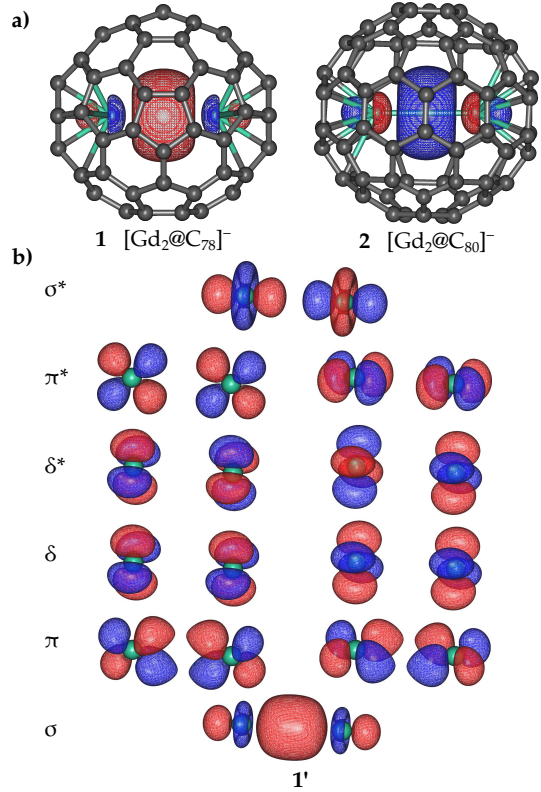}
  \caption{\label{F:orbitals} (a) The $\sigma$-bonding orbital housing the unpaired 5d electron in \textbf{1} and \textbf{2} and (b) the 5d orbital combinations in \textbf{1'} (the 5d orbital combinations in \textbf{2'} are very similar to the ones presented here).}
\end{figure}

The Kohn--Sham (KS) orbitals of \textbf{1} and \textbf{2} are largely similar to each other and the orbitals of \textbf{2} closely resemble those that have been reported earlier for the isoelectronic Gd$_2$@C$_{79}$N system.\cite{daul_2015a} The fourteen unpaired 4f electrons occupy seven bonding and seven anti-bonding combinations of the Gd 4f orbitals. These orbitals have only minor contributions from other Gd orbitals or from the cage orbitals and, thus, to a good approximation they can be considered as pure 4f combinations. The one ``extra'' electron (which will be from here on referred to as the 5d electron) occupies a $\sigma$-bonding type orbital with large amplitude in the Gd--Gd bonding region and some minor delocalization into the fullerene cage near the short Gd--C contacts (Figure \ref{F:orbitals}a).

The nature of the $\sigma$ orbital and the related Ln--Ln bonding interaction in analogous systems has been extensively discussed in the literature.\cite{popov_2012a,popov_2017a,shinohara_2014a} EPR measurements on related EMFs Y$_2$@C$_{79}$N and [La$_2$@C$_{80}$]$^{-}$ with empty 4f shells and one unpaired electron each are consistent with the unpaired spin being localized in the [Ln$_2$]$^{5+}$ core.\cite{dorn_2008a,kato_2007a} Furthermore, the magnitudes of the hyperfine coupling constants of the $^{139}$La and $^{89}$Y nuclear spins in the respective systems are relatively large. This suggests that the atomic s orbitals with large amplitudes near the nuclei make a significant contribution to the composition of the $\sigma$ bonding orbitals. In the present case, the $\sigma$ bonding orbitals of \textbf{1} and \textbf{2} consist mainly of 5d$_{\mathrm{z}^2}$, 6s and 6p$_\mathrm{z}$ orbitals. Based on L\"owdin reduced orbital populations, the relative contributions to the orbital composition are 27.0\% from d$_{\mathrm{z}^2}$ 24.6\% from s, 36.8\% from p$_\mathrm{z}$ and 26.6\% from d$_{\mathrm{z}^2}$ 20.8\% s, 32.6\% p$_\mathrm{z}$ for \textbf{1} and \textbf{2} respectively. In order to keep the notation simple and consistent with the description of the virtual orbitals (\emph{vide infra}), we will from here on refer to the $\sigma$ bonding orbital as a combination of the 5d$_{\mathrm{z}^2}$ orbitals while keeping in mind that in practice the orbitals are combinations of 5d$_{\mathrm{z}^2}$, 6s and 6p$_\mathrm{z}$ orbitals. Likewise, when we refer to the ``5d orbitals'' this should be interpreted as the $\pi$ and $\delta$ symmetric 5d orbitals and the bonding and anti-bonding $\sigma$ type orbitals.

The virtual 5d orbitals of \textbf{1} and \textbf{2} become extremely diffuse and strongly mix with the cage orbitals. In the dimer systems \textbf{1'} and \textbf{2'}, however, the virtual 5d orbital combinations can be easily identified and they are described in Figure \ref{F:orbitals}b. The bonding $\sigma$ symmetric orbitals of \textbf{1'} and \textbf{2'} show similar delocalization and mixing of s, p and d$_\mathrm{z^2}$ character as the respective orbitals in \textbf{1} and \textbf{2}. The $\pi$ and $\delta$ symmetric orbitals in \textbf{1'} and \textbf{2'} form two bonding and two anti-bonding combinations (labeled as $\pi^{*}$ and $\delta^{*}$) each . The $\pi$ orbitals show significant $\pi$-bonding character whereas the $\delta$ orbitals are largely non-bonding and confined to their respective Gd ions. Unlike the $\sigma$ bonding orbitals, the respective $\sigma^{*}$ orbitals consists almost purely of the Gd 5d$_{\mathrm{z}^2}$ orbitals.

Energies of the excited $\Sigma$, $\Pi$ and $\Delta$ terms obtained by 5d$\rightarrow$5d excitations in \textbf{1}, \textbf{2}, \textbf{1'} and \textbf{2'} were calculated with the DFT/ROCIS method and are listed in Table \ref{T:ROCIS}. For the dimers \textbf{1'} and \textbf{2'} the excitation energies corresponding to various 5d$\rightarrow$5d excitations can be easily identified by visual examination of the natural transition orbitals. In \textbf{1} and \textbf{2} the spectrum of 5d$\rightarrow$5d excitations becomes energetically mixed with the 5d$\rightarrow$cage and cage$\rightarrow$cage excitations and, therefore, the excited 5d configurations are strongly mixed with excited cage configurations. Because of this, only the lowest 5d$\rightarrow$5d excitations corresponding to $\sigma\rightarrow\pi$ excitations can be reliably assigned to a transition between two specific electronic configurations. The lowest excitation originating from the 4f shell lies at much higher energy than the $\sigma\rightarrow\sigma^{*}$ excitation in all systems considered.

\begin{table}
  \caption{5d$\rightarrow$5d excitation energies (in $\mathrm{cm}^{-1}$) calculated for \textbf{1}, \textbf{2}, \textbf{1'} and \textbf{2'} using the DFT/ROCIS method}
  \label{T:ROCIS}
  \begin{ruledtabular}
    \begin{tabular}{cdddd}
      Excitation & \textbf{1} & \textbf{2} & \textbf{1'} & \textbf{2'} \\
      \hline
      $\sigma\rightarrow\pi$         & 23293.2 & 21334.8 & 19357.2 & 21237.8 \\
      $\sigma\rightarrow\pi$         & 23331.5 & 22910.9 & 19357.2 & 21237.8 \\
      $\sigma\rightarrow\delta$      &         &         & 23630.4 & 26894.6 \\
      $\sigma\rightarrow\delta$      &         &         & 23630.5 & 26894.7 \\
      $\sigma\rightarrow\delta^{*}$  &         &         & 24177.2 & 27733.9 \\
      $\sigma\rightarrow\delta^{*}$  &         &         & 24177.3 & 27734.0 \\
      $\sigma\rightarrow\pi^{*}$     &         &         & 24577.2 & 28532.7 \\
      $\sigma\rightarrow\pi^{*}$     &         &         & 24577.2 & 28532.7 \\
      $\sigma\rightarrow\sigma^{*}$  &         &         & 25115.8 & 29441.0 \\
    \end{tabular}
  \end{ruledtabular}
\end{table}

The 5d$\rightarrow$5d excitation energies agree well with what can be qualitatively reasoned from the geometries and orbitals. The covalent interaction between the $\sigma$ symmetric orbitals in the ground state leads to strong stabilization of the ground configuration and therefore the first excited state lies roughly 20,000$\,\mathrm{cm}^{-1}$ above the ground state energy. The states in the manifold of excited 5d configurations lie much closer together. Under the $D_{3h}$ symmetry of \textbf{1} the $\Pi$ and $\Delta$ states should be exactly degenerate.  The small splitting of $38\,\mathrm{cm}^{-1}$ between the $\Pi$ states is most likely a result of the fact that the $D_{3h}$ symmetry was not explicitly imposed on the optimized wave functions. The (pseudo) $D_{2h}$ symmetry of \textbf{2} does not conserve any degeneracies and a splitting of $1576\,\mathrm{cm}^{-1}$ is observed in the $\Pi$ states.

\subsection{Magnetic properties of the ground configuration}

In their ground configuration \textbf{1} and \textbf{2} do not have any first order angular momentum because the lone 5d electron occupies an orbital of $\sigma$ symmetry which gives rise to a $\Sigma$ term. The orbital interaction is very strong and therefore electron delocalization governs the shape of the spectrum and splits the $\Sigma$ term into bonding and anti-bonding states. The bonding manifold is then further split by exchange interaction into states characterized by some total spin $S$. Thus, the magnetic interaction can be reduced to the isotropic exchange coupling between the two $S_{0,a}=S_{0,b}=S_0=7/2$ spins arising from the Gd 4f electrons and the $s=1/2$ of the 5d electron. The magnetic anisotropy, which emerges at second order, can be treated as a perturbation at a later stage.

The exchange splitting of the $\Sigma$ terms can be described as a ``classical'' three-site exchange coupled system where the Gd 4f electrons form two of the sites and the unpaired 5d electron forms the third site. The spin Hamiltonian for this system is written as 
\begin{equation}
  \label{E:ground_hamiltonian}
  \hat H_\text{3-site}
  = -J_\mathrm{Gd-Gd}\hat{\mathbf{S}}_{0,a}\cdot\hat{\mathbf{S}}_{0,b}
    - J_\mathrm{Gd-5d}\left(\hat{\mathbf{s}}\cdot\hat{\mathbf{S}}_{0,a} + \hat{\mathbf{s}}\cdot\hat{\mathbf{S}}_{0,b}\right)\text{,}
\end{equation}
where $\hat{\mathbf{S}}_{0,a}$ and $\hat{\mathbf{S}}_{0,b}$ act on the $S_0=7/2$ spins of the 4f electrons at Gd ion $a$ and $b$, respectively, $\hat{\mathbf{s}}$ acts on the spin of the 5d electron, and $J_\mathrm{Gd-Gd}$ and $J_\mathrm{Gd-5d}$ are the exchange coupling constants. 

An alternative approach would be to treat the system as a Gd(II)/Gd(III) mixed valence system with a 5d electron resonating between the two Gd ions. This model has been widely used, for example, in the description of Fe(II)/Fe(III) mixed valence complexes.\cite{girerd_1990a} However, the 5d electron in this model, although highly delocalized, would only have significant amplitude at atomic-like orbitals at the Gd ions and the model would then fail to describe the stabilization of the non-Hund states due to the delocalization of the electron into the bonding region.  Therefore, for the description of the ground states, we use the three-site model based on a delocalized $\sigma$ type orbital defined in the Hamiltonian (\ref{E:ground_hamiltonian}). The excited states, where the delocalization due to covalency is greatly reduced because of the much smaller overlap of the 5d orbitals, will, however, be interpreted in this manner (see Sec. \ref{S:pi_delta}).

In order to evaluate the spectrum of the exchange manifold of the ground configuration, the exchange coupling constants must be determined first. This can be achieved at DFT level using the broken symmetry (BS) formalism pioneered by Noodleman.\cite{noodleman_1981a,noodleman_1985a,noodleman_1986a} Values of the exchange coupling constants extracted from BS DFT calculations are $J_\mathrm{Gd-Gd} = -1.5\,\mathrm{cm}^{-1}$, $J_\mathrm{Gd-5d} = 388.6\,\mathrm{cm}^{-1}$ and $J_\mathrm{Gd-Gd} = -1.3\,\mathrm{cm}^{-1}$, $J_\mathrm{Gd-5d} = 354.2\,\mathrm{cm}^{-1}$ for \textbf{1} and \textbf{2}, respectively. The coupling constants are similar in magnitude to those calculated for the analogous Gd$_2$@C$_{79}$N system.\cite{daul_2015a} Details of the extraction procedure are given in Sec. II B. in the supplementary material.

Using the coupling constants obtained with BS DFT calculations and the eigenvalues of Eq. (\ref{E:ground_hamiltonian}) calculated in Sec. II A in the supplementary material, the exchange manifolds of the ground configurations of \textbf{1} and \textbf{2} (i.\,e. $\Sigma$ terms) were constructed and the resulting energy spectrum is presented in Figure \ref{F:exchange_states} (numerical values are available in Table S1 in the supplementary material). The low-lying spectrum of the exchange states was also calculated at CASSCF level for both \textbf{1} and \textbf{2} and the results (Table S1 in the supplementary material) agree qualitatively with the BS DFT results. CASPT2 calculations on \textbf{1} and \textbf{2} were not possible due to high computational costs and therefore the CASSCF results do not include any dynamic electron correlation. The exchange coupling constants extracted from BS DFT calculations do include dynamic correlation effects, although in an approximate manner. The spectrum constructed from BS DFT results should therefore, in principle, be more accurate than the CASSCF energies and all further discussion will be based on the BS DFT results.

The splitting between the average energies of the Hund and non-Hund manifolds is $1594\,\mathrm{cm}^{-1}$ and $1749\,\mathrm{cm}^{-1}$ for \textbf{1} and \textbf{2} respectively. The same splitting calculated for a single Gd(II) ion (see Table \ref{T:free_ion}) is $9025\,\mathrm{cm}^{-1}$. Thus, there is a significant reduction of the 4f--5d Hund's rule coupling strength in \textbf{1} and \textbf{2} as compared to the free ions. This can be explained by the mixing of the 6s and 6p$_\mathrm{z}$ orbitals with the 5d$_{\mathrm{z}^2}$ orbitals due to the lowering of the symmetry by the fullerene cage. The mixing leads to significant delocalization of the $\sigma$ symmetric orbital, as discussed earlier, and to a reduction in the coupling between the unpaired spin in the highly contracted 4f shell and the 5d electron.

\begin{figure}
  \includegraphics[width=7cm]{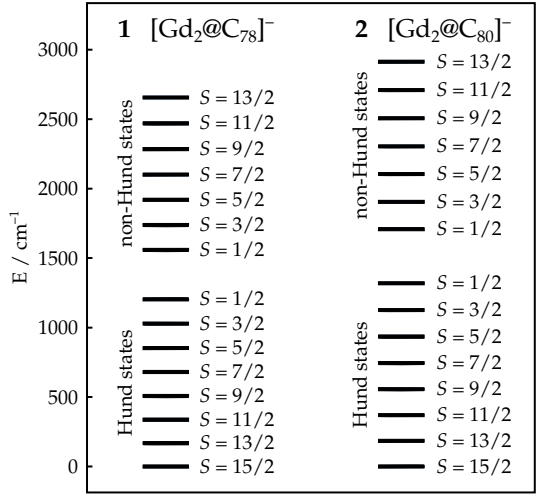}
  \caption{\label{F:exchange_states}Energies of the exchange states of the $\sigma$-bonding manifolds of \textbf{1} and \textbf{2} as calculated using Hamiltonian (\ref{E:ground_hamiltonian}) and exchange coupling constants extracted from broken-symmetry DFT calculations.}
\end{figure}

Zero-field splitting (ZFS) in the $S = 15/2$ ground state is weak due to the lack of first order angular momentum in the $\Sigma$ terms but not completely negligible because of the weak mixing of excited configurations into the ground configuration at higher orders of perturbation theory. The experimentally observed EPR spectra measured at $6\,\mathrm{K}$ is consistent with an $S = 15/2$ spin-state which is weakly split at zero field. The splitting pattern has been modeled by a giant spin Hamiltonian acting on the full $S$ multiplet affording the axial and rhobmic ZFS parameters $D$ and $E$, respectively.\cite{kato_2016a} The calculated splitting between the ground $\Sigma$ state and the first excited exchange state in \textbf{1} and \textbf{2} are $168\,\mathrm{cm}^{-1}$ and $184\,\mathrm{cm}^{-1}$, respectively, whereas the experimental ZFS parameters (see Table \ref{T:ZFS}) predict the splitting in the ZFS manifolds to be in the range of a few wave numbers. This means that the giant-spin approximation to the ZFS should be well-justified here and, considering the low temperature of the experimental conditions, thermal population of the $S=13/2$ exchange states can be safely neglected.

The ZFS parameters were first calculated at DFT level using the approach proposed by Neese\cite{neese_2006a,neese_2007b} and the values along with the experimentally determined parameters\cite{kato_2016a} are listed in Table \ref{T:ZFS}. The DFT results predict a negative sign for the $D$ parameter in both \textbf{1} and \textbf{2} which is opposite to what is experimentally observed. The negative sign would lead to a completely different electronic structure of the $S=15/2$ multiplet than what is experimentally observed. Therefore, the DFT results are clearly incorrect even at a qualitative level. This is not surprising considering that the DFT methods for the treatment of ZFS are known to give highly inaccurate results in a number of cases.\cite{neese_2015a,odelius_2013a,odelius_2015a} The results in Table \ref{T:ZFS} were calculated with the pure PBE GGA functional which does not include any exact exchange. Calculations with the hybrid PBE0 functional were also tried but this lead to divergence in the couple-perturbed equations. Other DFT based methods\cite{khanna_1999a,van-wullen_2011a} for the calculations of ZFS parameters where also attempted but did not lead to any visible improvement.

DFT calculation of ZFS parameters allows one to separate the contributions from SOC and spin-spin coupling (SSC). In case of lanthanide ions, both SOC and SSC can make sizable contribution to the ZFS tensor. Whereas the SOC contribution is extremely difficult to calculate at DFT level, the SSC contributions can be assumed to be more reliable. In the present case the SSC contributions to the $D$ and $E$ parameters are $D_{SS} = -0.042\,\mathrm{cm}^{-1}$, $E_{SS} = 0.000\,\mathrm{cm}^{-1}$ and $D_{SS} = 0.012\,\mathrm{cm}^{-1}$, $E_{SS} = 0.003\,\mathrm{cm}^{-1}$ for \textbf{1} and \textbf{2}, respectively. The magnitudes of the SSC parameters are therefore comparable to the SOC contributions. We note in passing that the ZFS parameters calculated purely from the SSC contributions are closer to the experiment than the combined SOC and SSC values but there is no theoretical justification for the neglection of the SOC contribution, and even if that was the case the sign of the $D_{SS}$ parameter in \textbf{1} would still be incorrect.

More accurate estimates of the ZFS parameters can, in principle, be obtained by \emph{ab initio} multireference calculations. However, the ZFS arises from the mixing of a large number of excited states into the ground spin multiplet due to the SOC and these states cannot be generated within some small orbital space outside the minimal CAS used in the CASSCF calculations. Meaningful results could only be obtained by considerably increasing the size of the CAS which would then render the calculations computationally untractable. A middle ground between the conventional DFT based methods for the calculations of ZFS parameters and the multireference methods can be reached in the DFT/ROCIS approach which includes SOC in a similar manner to the multireference calculations although using orbitals and orbital energies obtained from a restricted open-shell DFT calculations and some empirical scaling parameters.\cite{neese_2013b} A much larger number of high-lying excited states can be accounted with the DFT/ROCIS method as opposed to a CASSCF type calculation. Therefore, we also extracted the ZFS parameters from the energies and eigenvectors produced by the DFT/ROCIS calculation. Details of the extraction process are given in Sec. II C in the supplementary material and the values are listed in Table \ref{T:ZFS}. The magnitudes of the values of the $D$ parameters are still an order of magnitude larger than the experimental values but the sign is reproduced correctly. Considering the extremely small energy differences involved in the calculations and the fact that the energy differences are calculated from the full electronic energies, one is working at the limits of numerical accuracy and no quantitative agreement with experiment should be expected. The main observations made here should be that the sign of the $D$ parameters is positive and that the values are very small compared to the exchange splittings.

\begin{table}
  \caption{The calculated and experimental ZFS parameters of the ground $S=15/2$ multiplets of \textbf{1} and \textbf{2} in $\mathrm{cm}^{-1}$}
  \label{T:ZFS}
  \begin{ruledtabular}
    \begin{tabular}{ccrrr}
      && DFT & DFT/ROCIS& Experimental\cite{kato_2016a} \\
      \hline
      \textbf{1} & $D$ & $-0.233$ & $ 0.297$ & $0.0498$  \\
                 & $E$ & $ 0.000$ & $ 0.000$ & $0.00023$ \\
      \hline
      \textbf{2} & $D$ & $-0.099$ & $ 0.375$ & $0.0339$  \\
                 & $E$ & $-0.013$ & $ 0.017$ & $0.0102$  \\
    \end{tabular}
  \end{ruledtabular}
\end{table}

The ZFS parameters extracted from the DFT/ROCIS calculation reproduce, in addition to the experimentally observed sign of the $D$ parameter, the main experimentally observed difference between the anisotropies of \textbf{1} and \textbf{2}: the considerably larger rhombicity parameter $E$ in \textbf{2}. This can be rationalized based on symmetry considerations. The $D_{3h}$ symmetry of the geometry of \textbf{1} should lead to zero $E$ parameter and the calculated value is indeed exactly zero. The lower pseudo-$D_{2h}$ symmetry of \textbf{2} is not high enough to suppress the rhobmic components of the anisotropy. The experimentally observed $E$ in \textbf{1} is, however, very small but non-zero suggesting that under the experimental conditions \textbf{1} may show some very minor deviation from the exact $D_{3h}$ symmetry. It should be noted that the experimental values deviates from zero at the fourth decimal whereas the calculated parameters cannot be reliably distinguished from numerical noise beyond the third decimal and are therefore only three decimals are given in the reported values. The calculated $E$ parameter of \textbf{1} does indeed deviate from zero at the fourth decimal but it is impossible to say whether this results from deviation from the $D_{3h}$ symmetry (which has not been explicitly imposed on the wave function) or simply from numerical noise.

In addition to the $D$ and $E$ parameters which describe the second rank crystal field (CF) parameters, also higher order CF parameters up to rank 14 (which is the highest relevant rank for a $S=15/2$ system)  were extracted from the DFT/ROCIS calculation. These are listed in tables S2 and S3 in the supplementary material. In \textbf{1} the CF parameters of all ranks above the second are so small that they are either zero or they cannot be reliably distinguished from numerical noise. In the case of \textbf{2}, however, the fourth rank terms are still significant. This also becomes evident by constructing the energies of the zero-field split $S=15/2$ manifold by diagonalizing a CF Hamiltonian with various ranks of operators included (see table S4 and S5 in the supplementary material) and comparing these with the energies obtained by the DFT/ROCIS calculations. It is therefore not necessarily possible to exclude higher rank operators from the spin-Hamiltonian \emph{a priori} in systems such as \textbf{2}.

\subsection{Extraction of model Hamiltonian parameters}
\label{S:extraction}

In order to adequately describe the low-energy and excited electronic structures, we have derived the effective Hamiltonian describing the systems (\emph{vide infra}). The parameters used in the construction of the model Hamiltonians were extracted from various \emph{ab initio} and DFT calculations. The most important parameters governing the low-energy spectrum of the individual Gd(II) ions are the SOC constant $\zeta$ of the 5d electrons and the effective Hund's rule coupling parameter $J_H'$ which determines the energy difference between the Hund's states originating from the $^9D$ term and the non-Hund states originating from the $^7D$ term. The most important interionic parameters which determine the interaction energy between the two ions are the transfer integrals $t_{m_l}$ between 5d orbitals with orbital angular momentum projection $m_l$ on the two Gd ions.

The single-ion parameters were extracted from \emph{ab initio} calculations performed on a single Gd(II) ion. The \emph{ab initio} energies were compared with the eigenvalues of a model Hamiltonian expressed in terms of the parameters $\zeta$ and $J_H'$:
\begin{equation}
  \label{E:free_ion_hamiltonian}
  \hat H_\mathrm{Gd(II)} = \hat H_\mathrm{SOC} + \hat H_\mathrm{Hund} = \zeta\hat{\mathbf{l}}\cdot\hat{\mathbf{s}} - J_H'\hat{\mathbf{S}}_0\cdot\hat{\mathbf{s}}\text{,}
\end{equation}
where $\hat H_\mathrm{SOC}$ is the SOC Hamiltonian and $\hat H_\mathrm{Hund}$ is an effective Hund's rule coupling Hamiltonian which is used here in a Heisenberg-like form. \cite{anderson_1959a}  

The angular momentum operators $\hat{\mathbf{l}}$, $\hat{\mathbf{s}}$ and $\hat{\mathbf{S}}_0$ act on the orbital angular momentum of the 5d electron, the spin of the 5d electron and the total spin of the 4f electrons, respectively. 
Detailed derivation of the matrix elements of (\ref{E:free_ion_hamiltonian}) is presented in Sec. III A in the supplementary material.

Exact expressions of the eigenvalues of $\hat H_\mathrm{Gd(II)}$ are listed in Table \ref{T:free_ion}. Numerical values of $\zeta$ and $J_H'$ were extracted from the spectrum of $\hat H_\mathrm{Gd(II)}$ by performing a least squares fit of the energy differences between the ground state and a given excited state to the energy differences between the \emph{ab initio} calculated states. The fit yielded values $\zeta = 1037.84\,\mathrm{cm}^{-1}$ and $J_H' = 2069.53\,\mathrm{cm}^{-1}$. The relative errors compared to the \emph{ab initio} values are less than 10\% for all states and less than 3\% for more than half of the states. The energies calculated with these parameters, the \emph{ab initio} values, as well as experimentally determined values\cite{martin_1978} are also listed in Table \ref{T:free_ion}. The calculated values agree very well with the experiment.

\begin{table*}
  \caption{Energies of the spin-orbit coupled total angular momentum ($J$) states of a Gd(II) ion originating from the $^9D$ and $^7D$ terms as exact expressions obtained by a diagonalization of the model Hamiltonian (\ref{E:free_ion_hamiltonian}) and as numerical values obtained by fitting the exact expressions to \emph{ab initio} values as well as the \emph{ab initio} calculated and experimental values}
  \label{T:free_ion}
  \begin{ruledtabular}
    \begin{tabular}{ccrrr}
      State & Exact expression & Model / cm$^{-1}$ & \emph{Ab initio} / cm$^{-1}$ & Exp.\cite{martin_1978}/cm$^{-1}$ \\
      \hline
      $^9D_2$ & $\frac{1}{8} \left( - 13J_H' - 2\zeta - 2\sqrt{64 {J_H'}^2+64 J_H' \zeta+25 \zeta^2}\right)$ & $    0$ & $    0$ & $    0$ \\
      $^9D_3$ & $\frac{1}{8} \left( - 13J_H' - 2\zeta - 2\sqrt{64 {J_H'}^2+40 J_H' \zeta+25 \zeta^2}\right)$ & $  298$ & $  317$ & $  279$ \\
      $^9D_4$ & $\frac{1}{8} \left( - 13J_H' - 2\zeta - 2\sqrt{64 {J_H'}^2+8 J_H' \zeta+25 \zeta^2}\right) $ & $  758$ & $  775$ & $  694$ \\
      $^9D_5$ & $\frac{1}{8} \left( - 13J_H' - 2\zeta - 2\sqrt{64 {J_H'}^2-32 J_H' \zeta+25 \zeta^2}\right)$ & $ 1464$ & $ 1425$ & $ 1310$ \\
      $^9D_6$ & $\frac{1}{8} (-29J_H' + 8\zeta)                                                            $ & $ 2617$ & $ 2393$ & $ 2283$ \\
      $^7D_5$ & $\frac{1}{8} \left(-13J_H' - 2\zeta + 2\sqrt{64 {J_H'}^2-32 J_H' \zeta+25 \zeta^2}\right)  $ & $ 9404$ & $ 9045$ & $ 9356$ \\
      $^7D_4$ & $\frac{1}{8} \left(-13J_H' - 2\zeta + 2\sqrt{64 {J_H'}^2+8 J_H' \zeta+25 \zeta^2}\right)   $ & $ 9869$ & $ 9695$ & $ 9718$ \\
      $^7D_3$ & $\frac{1}{8} \left(-13J_H' - 2\zeta + 2\sqrt{64 {J_H'}^2+40 J_H' \zeta+25 \zeta^2}\right)  $ & $10137$ & $10153$ & $10015$ \\
      $^7D_2$ & $\frac{1}{8} \left(-13J_H' - 2\zeta + 2\sqrt{64 {J_H'}^2+64 J_H' \zeta+25 \zeta^2}\right)  $ & $10264$ & $10470$ & $10234$ \\
      $^7D_1$ & $\frac{1}{8} (3J + 8\zeta)                                                                 $ & $10372$ & $10672$ & $10387$ \\
    \end{tabular}
  \end{ruledtabular}
\end{table*}

The transfer parameters, $t_{m_l}$, were extracted from DFT calculations. First a set of Kohn--Sham (KS) orbitals of interest were localized onto the Gd ions so that they maximally resemble their atomic orbital counterparts while still retaining their polarization and hybridization due to the environment. Then, the KS Hamiltonian was transformed into this basis. Within the subspace of the localized orbitals, the transformed KS Hamiltonian has a one-to-one correspondence with a tight-binding Hamiltonian
\begin{align}
  \label{E:tight-binding}
  \hat{H}_\text{tb}
  = \sum_{\mu,m_s} \epsilon_\mu|a\mu m_s\rangle \langle a\mu m_s| 
  + \sum_{\nu,m_s} \epsilon_\nu|b\nu m_s\rangle \langle b\nu m_s| \nonumber \\
  + \sum_{\mu,\nu,m_s} t_{\mu \nu}\left(|a\mu m_s\rangle \langle b\nu m_s| + |b\nu m_s\rangle \langle a\mu m_s|\right),
\end{align}
where index $\mu$ $(\nu)$ runs over all 4f and 5d orbitals on ion $a$ ($b$), $\epsilon_\mu$ and $\epsilon_\nu$ are atomic orbital energies, $t_{\mu\nu}$ is a transfer parameter, and $m_s$ is the spin projection. The off-diagonal elements in the subspace of the localized orbitals are simply the transfer parameters and therefore the off-diagonal elements of the KS Fock operator in this basis can be identified as the transfer parameters of the tight-binding Hamiltonian.

The calculations were carried out for orbitals and eigenvalues obtained with both the hybrid PBE0 functional and the pure PBE GGA functional. The exact exchange in the PBE0 functional reduces the delocalization error compared to the pure PBE functional and should therefore offer more accurate results. However, the KS Fock operator constructed using the PBE0 potential includes a contribution from the Hatree--Fock exchange operator, and therefore the occupied and virtual orbitals do not feel the same potential. This means that, in the case of the PBE0 functional, localization of the $\sigma$ symmetric orbital which would require mixing of occupied and virtual canonical orbitals is not possible without introducing unphysical artifacts, and therefore the $\sigma\leftrightarrow\sigma$ transfer parameters are only available using the GGA functional.

The 4f orbital combinations are easy to identify, and their localization poses no considerable challenges. Unfortunately, in the case of \textbf{1} and \textbf{2} the virtual 5d orbitals become strongly mixed with the cage orbitals and isolating a set of 5d orbital combinations from the virtual orbital space is not possible without including an arbitrary number of cage, 6d, 6s, 6p etc. orbitals into this set. Localization of this arbitrary set would then lead to arbitrary values of transfer parameters which depend on the size of the chosen orbital set and cannot be determined in a unique way. To avoid this problem, the transfer parameters were extracted from the simple [Gd$_2$]$^{5+}$ dimers \textbf{1'} and \textbf{2'}. The effect of the cage on the direct interaction between the two Gd ions is assumed to be small and the transfer parameters extracted from calculations on \textbf{1'} and \textbf{2'} should therefore be a good approximation to the respective values in \textbf{1} and \textbf{2}. The calculated transfer parameters are listed in Table \ref{T:transfer} as determined with both the PBE0 and PBE functionals. Values extracted from DFT/ROCIS calculations in section \ref{S:splitting} are listed in the same table. Complete list of the 5d$\leftrightarrow$5d and 4f$\leftrightarrow$4f parameters calculated at PBE0 level is given in tables S6 and S7 in the supplementary material.

The main consequence resulting from neglection of the cage is that the symmetry of both \textbf{1'} and \textbf{2'} is strictly axial and therefore only transfer parameters between 5d orbitals corresponding to the same value of $m_l$ on the two ions have non-zero values. The values determined with the PBE0 and PBE functionals are very similar to each other with the PBE values being slightly smaller in magnitude. In all subsequent calculations the PBE0 values will be used but considering the similarity of the values either set should produce comparable results. The explicit value of the $\sigma\leftrightarrow\sigma$ parameter is not needed in any of the calculations beyond knowing that it is larger than the two other parameters.

\begin{table*}
  \caption{Transfer parameters between 5d orbitals of the two Gd ions in \textbf{1'} and \textbf{2'} as extracted from DFT calculations using the PBE0 and PBE functionals and as extracted from DFT/ROCIS calculations}
  \label{T:transfer}
  \begin{ruledtabular}
    \begin{tabular}{ccccccc}
      & \multicolumn{2}{c}{PBE0} & \multicolumn{2}{c}{PBE} & \multicolumn{2}{c}{DFT/ROCIS} \\
      & $t$ (\textbf{1'}) / cm$^{-1}$  & $t$ (\textbf{2'}) / cm$^{-1}$
      & $t$ (\textbf{1'}) / cm$^{-1}$  & $t$ (\textbf{2'}) / cm$^{-1}$
      & $t$ (\textbf{1'}) / cm$^{-1}$  & $t$ (\textbf{2'}) / cm$^{-1}$ \\
      \hline
      $\sigma\leftrightarrow\sigma$ &            &            &  $9104.2$ & $11125.4$ & $12557.9$ & $14720.5$ \\
      $\pi\leftrightarrow\pi$       & $  2667.8$ & $  3715.0$ &  $2275.6$ &  $3223.1$ & $ 2610.0$ & $ 3647.5$ \\
      $\pi\leftrightarrow\pi$       & $  2667.8$ & $  3715.0$ &  $2275.6$ &  $3223.1$ & $ 2610.0$ & $ 3647.5$ \\
      $\delta\leftrightarrow\delta$ & $   268.1$ & $   412.8$ &   $231.6$ &   $358.4$ & $  273.4$ & $  419.7$ \\
      $\delta\leftrightarrow\delta$ & $   268.7$ & $   412.8$ &   $231.5$ &   $358.6$ & $  273.4$ & $  419.7$ \\
    \end{tabular}
  \end{ruledtabular}
\end{table*}

\subsection{Splitting between $\Sigma$, $\Pi$ and $\Delta$ terms}
\label{S:splitting}

Before we discuss the splitting of the $\Pi$ and $\Delta$ terms under the influence of Hund's rule coupling, SOC and electron transfer in the next section, we must first determine the relative energies of the energy manifolds arising from the $\Sigma$, $\Pi$ and $\Delta$ terms. In the $S=15/2$ high-spin state, the splitting of the single-ion energy levels due to electron transfer is $\pm t_{m_l}$ and therefore the difference between the bonding and anti-bonding states is simply $2t_{m_l}$.\cite{kahn_1993} Therefore, energies of the free-ions terms can simply be calculated as the middle point in energy between the respective bonding and anti-bonding states. The transfer parameters extracted from the DFT/ROCIS energy differences are listed in Table \ref{T:transfer} along with the values determined in the previous section. It is clear that the PBE0 and the DFT/ROCIS transfer parameters for the $\pi$ and $\delta$ orbitals are very similar as should be expected.

Setting the energy origin at the $S=15/2$ spin-state of the free-ion $\Sigma$ term, the energies of the crystal-field split free-ion $\Pi$ and $\Delta$ terms in their $S=15/2$ spin-state are $E_\Pi=9409\,\mathrm{cm}^{-1}$, $E_\Delta=11346\,\mathrm{cm}^{-1}$ and $E_\Pi=10165\,\mathrm{cm}^{-1}$, $E_\Delta=12594\,\mathrm{cm}^{-1}$ for \textbf{1} and \textbf{2} respectively.

\subsection{Energy spectrum of the excited $\pi$ and $\delta$ configurations}
\label{S:pi_delta}

Magnetic properties of the excited configurations of \textbf{1} and \textbf{2}, where the lone 5d electron is promoted to the $\pi$ or $\delta$ orbitals and which give rise to $\Pi$ and $\Delta$ terms, are very different from those of the ground configuration. Firstly, the overlap between the $\pi$ and $\delta$ symmetric 5d orbitals on the two ions is much smaller than that between the $\sigma$ symmetric orbitals. This leads to much weaker covalent bonding and thus more localized orbitals and smaller magnitudes of transfer parameters. Secondly, the orbital angular momentum of the $\pi$ or $\delta$ orbitals is not quenched and first order angular momentum enters the equations. The SOC constant determined earlier for the Gd(II) ions is $\zeta = 1037.84\,\mathrm{cm}^{-1}$ which is roughly of the same order of magnitude as the effective Hund's rule coupling constant $J_H' = 2069.53\,\mathrm{cm}^{-1}$ and the transfer parameters of the $\pi$ or $\delta$ orbitals listed in Table \ref{T:transfer}. Therefore, all of these interactions must be treated on equivalent footing. The exchange interaction between the 4f electrons (described by $J_\mathrm{Gd-Gd}$ in the ground configuration) is, however, several orders of magnitude smaller than the other interactions and will be neglected in all subsequent calculations. Under these conditions, the excited states can be viewed as a mixed valence Gd(II)/Gd(III) system and the magnetism can be described in terms of a double exchange model. The general idea of the model is that the 5d electron resonates between the two sites where it is coupled to the 4f spins. This introduces a spin-dependent delocalization into the system. The concept  was originally proposed in the context of ionic solids by Zener\cite{zener_1951a}, Hasegawa and Anderson\cite{anderson_1955a} and de Gennes\cite{de-gennes_1960a}, and later adapted to molecular systems by Girerd\cite{girerd_1983a,girerd_1987a,girerd_1990a} and Noodleman\cite{noodleman_1984a}. Unlike in a conventional treatment of the double exchange, however, the presence of unquenched first-order orbital angular momentum in the present case means that SOC has to be explicitly introduced into the model and the splitting of energy levels will be very different from the spin-only case.

The crystal-field removes the five-fold degeneracy of the orbital $l=2$ states. We will assume a crystal-field with an axial symmetry (trigonal or higher) and, thus, the crystal-field will retain the two-fold orbital degeneracies of the $\Pi$ and $\Delta$ states. In case of \textbf{1} this is correct but for \textbf{2} this assumption constitutes an approximation. We will base most of our discussion on the strong crystal-field limit where the crystal-field splitting is assumed strong enough so that mixing of the $m_l=0$ state into the $m_l=\pm 1$ states and $m_l=\pm 1$ into $m_l=\pm 2$ states by SOC can be neglected. The splitting between the $\Pi$ and $\Delta$ states determined in section \ref{S:splitting} are $1937\,\mathrm{cm}^{-1}$ and $2429\,\mathrm{cm}^{-1}$ for \textbf{1} and \textbf{2}, respectively, which is roughly twice the SOC constant and therefore neglecting the $\Pi-\Delta$ mixing is undoubtedly an approximation. The effect of this simplification to the results derived in this section will be discussed later in section \ref{S:mixing} and it will be shown that the conclusions made here remain valid even when the mixing is taken into account.

The full Hamiltonian within a given $m_l=\pm 1$ or $m_l\pm 2$ crystal-field doublet is of the form
\begin{align}
  \hat H_\mathrm{full} = \hat H_\mathrm{SOC}^a + \hat H_\mathrm{Hund}^a + \hat H_\mathrm{SOC}^b + \hat H_\mathrm{Hund}^b + \hat H_\mathrm{transfer}\text{,}
\end{align}
where the superscripts $a$ and $b$ indicate operators that only act on states where the 5d electron is localized at ion $a$ or $b$ and $\hat H_\mathrm{transfer}$ is the transfer Hamiltonian which couples the localized states. We will first diagonalize the single-ion Hamiltonians ($\hat H_\mathrm{SOC}^a + \hat H_\mathrm{Hund}^a$ and $\hat H_\mathrm{SOC}^b + \hat H_\mathrm{Hund}^b$ for ions $a$ and $b$ respectively) to account for SOC and Hund's rule coupling and then consider the resonance stabilization of these states due to the electron delocalization.

Derivation of the matrix elements of the single-ion Hamiltonians and details of the diagonalization procedure are given in Sec. III B in the supplementary material. The eigenvalues are
\begin{align}
  \label{E:eigenvalues}
  E_{\pm}&(J_H',\zeta;S_0,M_K,m_l) \\ \nonumber
  & = \frac{1}{2}J_H'
  \pm\frac{1}{2}\left[J_H'^2s^2(2S_0+1)^2 - 8J_H'\zeta m_ls^2M_K \phantom{\frac{16s^2(S_0+s)}{(2S_0+1)^2}}\right. \\ \nonumber
  &\qquad\qquad\qquad\qquad + \left.\zeta^2m_l^2\frac{16s^2(S_0+s)}{(2S_0+1)^2}\right]^{1/2}
\end{align}
where we have only considered the case where $s=1/2$ and $S_{0,a}=S_{0,b}\equiv S_0$. The middle term in the square brackets ensure that the states with the same absolute total single-ion angular momentum projection $|M_K+m_l|$ are degenerate. This introduces a two-fold degeneracy in the states. The single-ion energies do not depend on the orientation of the 4f spins of the uncoupled ion which introduced an additional eight-fold degeneracy due to the eight possible projections of the $S_{0,a}=7/2$ or $S_{0,b}=7/2$ spins of the other ion. Therefore, the total degeneracy of the single-ion eigenstates is 16. These degeneracies cannot be completely lifted by $\hat H_\mathrm{transfer}$ as at least double degeneracy must be retained in all states due to Kramers' theorem. 

\begin{figure}
  \includegraphics[width=7cm]{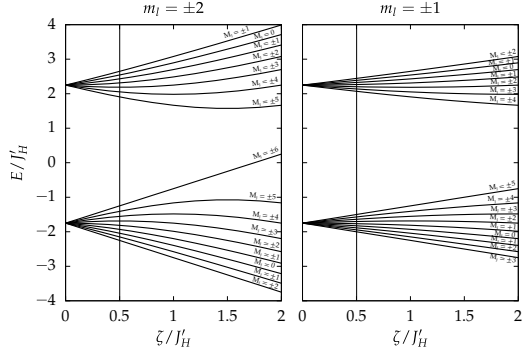}
  \caption{\label{F:single_ion_states}Effect of the $\zeta/J_H'$ ratio on the splitting of single-ion energy levels within a $m_l=\pm 2$ and $m_l=\pm 1$ crystal field doublets. The vertical lines indicate the $\zeta/J_H'$ ratio calculated with the \emph{ab initio} values.}
\end{figure}

Numerical values of $E_{\pm}(J_H',\zeta;S_0,M_K,m_l)$ are listed in Table \ref{T:single-ion_energies}. The eigenvectors are linear combinations of a Hund and a non-Hund state corresponding to the same value of $M_K$ except for $M_K=4$ and $M_K=-4$ when there is only one available basis state and the eigenvector consists purely of this state. Splitting of the single-ion energy levels as a function of the $\zeta/J_H'$ ratio is presented in Figure \ref{F:single_ion_states} for the $m_l=\pm 1$ and $m_l=\pm 2$ crystal-field doublets. At zero $\zeta$ the Hund and non-Hund states form two degenerate manifolds which are split once the value of the $\zeta/J_H'$ ratio is increased. When $\zeta/J_H'\lesssim 1$ SOC causes a linear splitting within the manifolds and as the ratio increases further the Hund and non-Hund manifolds become increasingly mixed. At small values of $\zeta/J_H'$, the lower manifold consists of nine energy states and the higher manifold consists of seven energy states but once $\zeta/J_H'\gtrsim 1$ the $M_J=\pm 5$ (in the case of the $m_l=\pm 1$ crystal-field doublet) and the $M_J=\pm 6$ (in the case of the $m_l=\pm 2$ doublet) states linearly transfer from the lower manifold to the higher-energy manifold. When the $\zeta/J_H'$ ratio is further increased the splitting between the two manifolds tends towards infinity. No energy level crossing take place at any values of $\zeta/J_H'$. It should be noted that in the real physical situation, increasing $\zeta$ will also lead to mixing of the $m_l=\pm 1$ and $m_l=\pm 2$ crystal-field doublets as will be discussed in section \ref{S:mixing}. Eventually SOC will become of similar magnitude and stronger than the crystal-field splitting and the picture presented in Figure \ref{F:single_ion_states} will break down. In the $m_l=\pm 1$ crystal-field doublet $M_J=\pm 3$ states lie lowest in energy and in the case of the $m_l=\pm 2$ doublet, $M_J=\pm 2$ energy states are the lowest. If $J_H'\rightarrow\infty$ (i.e. the coupling between the Hund and non-Hund states can be neglected) the energy of the different $M_K$ states in the Hund manifold can be expressed as (see Sec. III B in the supplementary material):
\begin{align}
  E_{J_H'\rightarrow\infty}(\zeta;M_K,m_l) = \frac{2\zeta m_lsM_K}{2S_0+1}
\end{align}
which clearly shows the linear splitting of different $M_K$ states as a function of $\zeta$ observed in Figure \ref{F:single_ion_states} at small values of $\zeta/J_H'$ ratio when the mixing of the Hund and non-Hund manifolds is negligible.

\begin{table}
  \caption{Numerical values of the energies $E_{\pm}(J_H',\zeta;S_0,M_K,m_l)$ (in $\mathrm{cm}^{-1}$) of the single-ion Hamiltonians calculated for the $m_l=\pm 1$ and $m_l=\pm 2$ crystal-field doublets.}
  \label{T:single-ion_energies}
  \begin{ruledtabular}
    \begin{tabular}{rccrc}
      $M_J$ & $E(m_l = \pm 1)$ & & $M_J$ & $E(m_l = \pm 2)$ \\
      \hline
      $\pm 3$ & $   0$ & & $\pm 2$ & $    0$ \\
      $\pm 2$ & $ 117$ & & $\pm 1$ & $  212$ \\
      $\pm 1$ & $ 237$ & & $\pm 0$ & $  433$ \\
      $\pm 0$ & $ 360$ & & $\pm 1$ & $  665$ \\
      $\pm 1$ & $ 487$ & & $\pm 2$ & $  910$ \\
      $\pm 2$ & $ 617$ & & $\pm 3$ & $ 1169$ \\
      $\pm 3$ & $ 752$ & & $\pm 4$ & $ 1447$ \\
      $\pm 4$ & $ 892$ & & $\pm 5$ & $ 1747$ \\
      $\pm 5$ & $1038$ & & $\pm 6$ & $ 2076$ \\
      $\pm 4$ & $8423$ & & $\pm 5$ & $ 8607$ \\    
      $\pm 3$ & $8564$ & & $\pm 4$ & $ 8907$ \\
      $\pm 2$ & $8699$ & & $\pm 3$ & $ 9185$ \\    
      $\pm 1$ & $8829$ & & $\pm 2$ & $ 9444$ \\
      $\pm 0$ & $8956$ & & $\pm 1$ & $ 9689$ \\
      $\pm 1$ & $9079$ & & $\pm 0$ & $ 9921$ \\    
      $\pm 2$ & $9199$ & & $\pm 1$ & $10142$ \\
    \end{tabular}
  \end{ruledtabular}
\end{table}

\begin{figure}
  \includegraphics[width=0.90\linewidth]{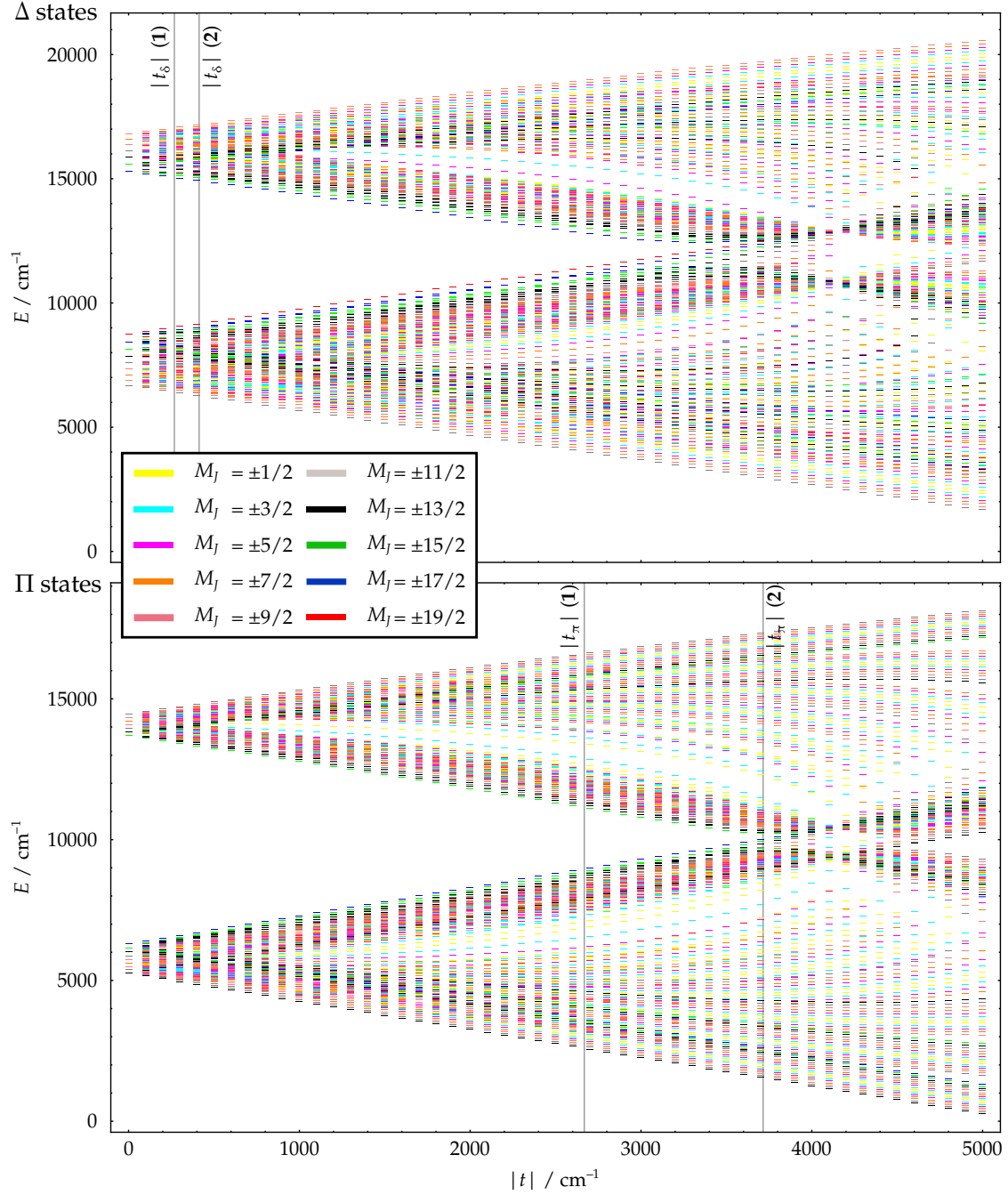}
  \caption{\label{F:transfer_splitting}Splitting of the $\Pi$ (bottom) and $\Delta$ (top) single-ions states under the influence of electron transfer in \textbf{1} and \textbf{2} as a function of the magnitude of the transfer parameter. The energy levels are color-coded based on the  projection of the total angular momentum. Vertical lines indicate the calculated values of the transfer parameters. The energy scale on the vertical axis is that of \textbf{1}; in case of \textbf{2} all energy levels are translated to a slightly higher energy due to the larger crystal-field splitting.}
\end{figure}

The transfer Hamiltonian $\hat H_\mathrm{transfer}$ mixing the single-ion states is of the same form as in the tight-binding Hamiltonian (\ref{E:tight-binding}). 
Derivation of its matrix elements is given in Sec. III C in the supplementary material. Due to the assumed axial symmetry, $\hat{H}_\text{transfer}$ conserves the total angular momentum projection and the value of $m_l$. Therefore, the $\hat H_\mathrm{full}$ matrix will be block-diagonal in blocks corresponding to the same values of $M_K+M_{0,b}$ or $M_K'+M_{0,a}'$ and $m_l$. Although the smallest blocks can be easily diagonalized analytically, the characteristic polynomials of the larger blocks cannot be solved exactly to yield analytical expressions for the eigenvalues. Thus, the energy spectrum of $\hat{H}_\text{full}$ was obtained by numerical diagonalization. $\hat H_\mathrm{full}$ has 128 unique eigenvalues for both crystal-field doublets, each of which is doubly degenerate. The splitting of the $\Pi$ and $\Delta$ single-ions states under the influence of electron transfer is presented in Figure \ref{F:transfer_splitting} as a function of the respective transfer parameter.

It is immediately clear from Figure \ref{F:transfer_splitting} that the single-ion $\Pi$ and $\Delta$ states become energetically mixed even at relatively small values of $|t|$. In the range $|t|\lesssim 2500\,\mathrm{cm}^{-1}$ the Hund and non-Hund manifolds of the $\Pi$ and $\Delta$ states are clearly separated from each other. In the same range of values the splitting of the states into bonding and anti-bonding manifolds is also observable in the spectrum with the highest density of states in the top and bottom of the Hund and non-Hund manifolds. Beyond $|t|\sim 3000\,\mathrm{cm}^{-1}$ the interaction approaches that of a covalently bound system and the states in the Hund and non-Hund manifolds with the same value of $M_J$ become mixed and all manifolds become energetically intertwined. Considering the $\Delta$ state energy spectrum calculated using the $\delta\leftrightarrow\delta$ transfer parameters of \textbf{1} and \textbf{2}, the splitting between the Hund and non-Hund manifolds is larger than the transfer-splitting within these manifolds. In the absence of SOC this situation could be described by the conventional double exchange mechanism. On the other hand, the spectrum of the $\Pi$ states calculated using the $\pi\leftrightarrow\pi$ transfer parameters of \textbf{1} and \textbf{2} displays two different situations. In the case of \textbf{1}, the magnitude of the transfer parameter is such that the transfer-splitting is larger than the separation between the Hund and non-Hund manifolds but still small enough to retain a clear splitting between the two manifolds. The transfer parameter of \textbf{2} is, however, already so large that the interaction is better described as weak covalent interaction than double exchange.

The strong mixing of single-ion states by the transfer interaction makes analysis of the spectrum difficult and it is therefore instructive to consider the case when $J_H'\rightarrow\infty$ and $|t|\ll\zeta$ where an approximate analytical form can be given for the eigenvalues. Under these conditions the interaction between single-ion states corresponding to different values of $M_K$ can be neglected and conservation of the angular momentum projection under transfer interaction then implies that $M_{0,b}=M_{0,a}'\equiv M_0$ and $m_l=m_l'$. Eigenvalues within a manifold of states corresponding to a given value of $M_K$ (see Sec. III C in the supplementary material) is
\begin{align}
  \label{E:small_t_splitting}
  E_\pm&(t;K,M_K,M_0,m_l) \\ \nonumber
  & = \left\{
  \begin{array}{ll}
    \dfrac{2\zeta m_lsM_K \pm t\left(K + M_K\right)}{2S_0+1} & \text{if } M_0=M_K-s \\
    \dfrac{2\zeta m_lsM_K \pm t\left(K - M_K\right)}{2S_0+1} & \text{if } M_0=M_K+s \\    
    \dfrac{2\zeta m_lsM_K}{2S_0+1}                             & \text{else.}
  \end{array}\right.
\end{align}
In the case $M_J=\pm 19/2$ the expression is exact and in other cases it is equivalent to the first order perturbation correction to the single-ion energies due to transfer interaction. Figures S2 and S3 in the supplementary material show the splitting of $\Delta$ states in the Hund manifold of \textbf{1} according to (\ref{E:small_t_splitting}) as a function of $|t|$ both in the cases when the coupling between the Hund and non-Hund manifolds is included and when it is neglected. The results show that for large values of $|M_J|$ equation (\ref{E:small_t_splitting}) describes the splitting reasonably well in the $0<|t|<100\,\mathrm{cm}^{-1}$ range but for smaller values of $|M_J|$ it is only qualitatively correct.

Equation (\ref{E:small_t_splitting}) and the figures S2 and S3 show that for a given value of $|t|$ (assuming $|t|\ll\zeta$) the transfer splitting is linearly proportional both to $K/(2S_0+1)$ and to $M_K/(2S_0+1)$. This is in sharp contrast to the splitting due to the conventional isotropic double exchange mechanism where the splitting is linearly proportional to $(S+s)/(2S_0+1)$ where $S$ is the total spin of the system.\cite{kahn_1993} Therefore, in a system with an axial symmetry, the presence of strong anisotropy introduces an Ising-like dependence on the magnitude of the transfer splitting at the strong crystal-field limit.

\subsection{Mixing of states arising from the $\pi$ and $\delta$ configurations}
\label{S:mixing}

The crystal-field splitting between the $\Pi$ and $\Delta$ states is, as determined in section \ref{S:splitting}, $1937\,\mathrm{cm}^{-1}$ for \textbf{1} and $2429\,\mathrm{cm}^{-1}$ for \textbf{1}. In both cases this is roughly twice the size of the SOC constant ($\zeta=1037.84\,\mathrm{cm}^{-1}$) and therefore some mixing between the $\Pi$ and $\Delta$ terms due to SOC is to be expected. It is, thus, relevant to discuss to what extent this mixing affects the results derived in the previous section.

\begin{figure}
  \includegraphics[width=7cm]{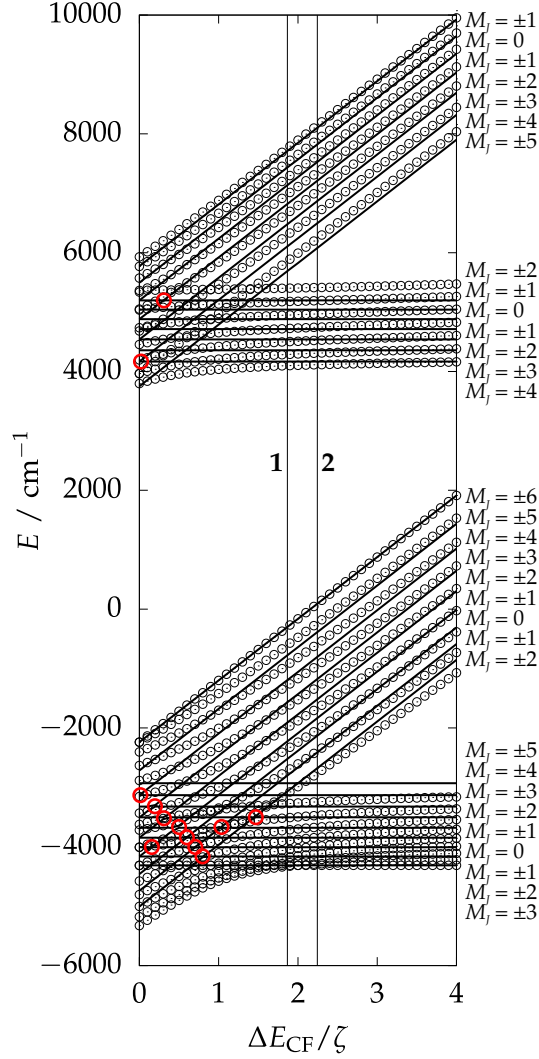}
  \caption{\label{F:CF_coupling} The single-ion energy levels as a function of the crystal-field splitting between the $\Pi$ and $\Delta$ states ($\Delta E_\mathrm{CF}$). The solid line describe the splitting when the mixing between $\Pi$ and $\Delta$ states is neglected (equation (\ref{E:eigenvalues})) and black circles indicate the values as calculated by numerical diagonalizition of the single-ion Hamiltonian (\ref{E:single_ion}) with the $\Pi$--$\Delta$ mixing included. Red circles indicate crossings of energy levels (as calculated with (\ref{E:eigenvalues}))  with the same value of $M_J$. Vertical lines indicate the values of crystal-field splitting as calculated for \textbf{1} and \textbf{2} at DFT/ROCIS level.}
\end{figure}

The single-ion Hamiltonian acting in the basis of both the $\Pi$ and $\Delta$ manifolds reads
\begin{equation}
  \label{E:single_ion}
  \hat H_\mathrm{single-ion}
  = \hat H_\mathrm{SOC} + \hat H_\mathrm{Hund} + \hat H_{\mathrm{CF},\Delta}\text{,}
\end{equation}
where $\hat H_{\mathrm{CF},\Delta}$ simply adds the crystal-field splitting energy $\Delta E_\mathrm{CF}$ to the diagonal elements of the $m_l=\pm 2$ states. The matrix elements of $\hat H_\mathrm{SOC}$ and $\hat H_\mathrm{Hund}$ can be calculated as derived in Sec. III A of the supplementary material. The characteristic polynomials of the matrix cannot be solved analytically and the matrix can only be diagonalized numerically. The eigenvalues of $\hat H_\mathrm{single-ion}$ as a function of $\Delta E_\mathrm{CF}$ are presented in Figure \ref{F:CF_coupling} along with the eigenvalues calculated using equation (\ref{E:eigenvalues}). In the range $E_\mathrm{CF}<1.5\zeta$ the mixing is very strong. Beyond this value the spectrum is qualitatively similar to that calculated with equation (\ref{E:eigenvalues}) with the eigenvalues being slightly shifted in energy. As $E_\mathrm{CF}$ is further increased, the approximate values slowly converge towards the exact eigenvalues.

Mixing of the $m_l=\pm 1$ states into the $m_l=\pm 2$ states means that $M_K$ and $m_l$ are no longer good quantum numbers. The single-ion states are still characterized by the total angular momentum projection $M_J$ and the double degeneracy of the $|M_J|$ states is retained as can be expected as Kramers degeneracy is not lifted. Therefore, the $\Pi$--$\Delta$ mixing does not break any degeneracies. All crossings of energy levels belonging to the same value of $|M_J|$ (marked by red circles in Figure \ref{F:CF_coupling}) take place at crystal-field splitting $E_\mathrm{CF}<1.5\zeta$. Therefore, all the qualitative changes in the energy level spectrum take place at crystal-field splittings weaker than those determined for \textbf{1} and \textbf{2} at DFT/ROCIS level ($1.87\zeta$ and $2.24\zeta$, respectively). Based on these consideration, it is safe to conclude that in the case of \textbf{1} and \textbf{2}, although the $\Pi$--$\Delta$ mixing inevitably introduces some error into the eigenvalues calculated in the previous section, all the qualitative features of the energy spectrum that can be deduced from the equations given in section \ref{S:pi_delta} remain valid even when the $\Pi$--$\Delta$ mixing is taken into account.

\subsection{Energy spectrum of the excited 4f configurations}

For the sake of completeness and with possible applications to other [Ln$_2$]$^{5+}$ systems in mind, we will also discuss the situation where the ``extra'' electron occupies a 4f orbital. In a free Gd(II) ion the states arising from the 4f$^8$ configuration are much higher in energy than the states arising from the 4f$^7$5d$^1$ configuration. Energy of the $^7F$ term before the inclusion of SOC is calculated at CASSCF(8,16)/XMS-CASPT2 level as $41,527\,\mathrm{cm}^{-1}$ compared to the ground state.

We will only consider the Hund configurations where the seven 4f electrons of a Gd(III) ion have the same spin and the electron resonating between the ions must then have a different spin. Due to the strong shielding of the 4f orbitals by 5s and 5p orbitals and the large SOC constant of the 4f orbitals, the splitting due to SOC is assumed much larger than the crystal-field splitting. Therefore we will first consider the coupling of the $l=3$ orbital angular momentum with the spin $K$ to give a total single-ion angular momentum $J_0$. According to Hund's rules (the 4f shell is more than half-filled) $J_0=K+l=6$. As before, we assume an axial crystal field. The crystal field then splits the $(2J_0+1)$-fold degenerate manifold of states corresponding to different projections $M_{J0}$ into 6 pairs of doublets characterized by non-zero $|M_{J0}|$ and one singlet with $M_{J0}=0$. When the ``extra'' electron is localized at ion $a$, this ion is described by $|aKlJ_0M_{J0}\rangle$ and the other ion $b$ by $|S_{0,b}M_{0,b}\rangle$, and hence, in the absence of transfer interaction the dimer is described by the direct product state $|KlJ_0M_{J0}\rangle\otimes|S_{0,b}M_{0,b}\rangle$. The opposite situation is expressed by exchanging the the indices $a$ and $b$. 

When $S_{0,a}=S_{0,b}\equiv S_0$ both direct product states are degenerate and each of them has a $(2S_0+1)$-fold degeneracy due to the different values of $M_{0,a}=M_{0,b}\equiv M_0$ thus raising the total degeneracy to $4S_0+2$. These states are then mixed by the transfer Hamiltonian $\hat H_\mathrm{transfer}$. Since the transfer parameters between the 4f orbitals are small (see Tables S6 and S7 in the supplementary material), whereas the crystal-field is strong due to the short Gd--Gd distance, we can neglect the mixing of the crystal-field states corresponding to different values of $|M_{J0}|$ by $\hat{H}_\text{transfer}$ to a good approximation. Calculation of the matrix elements and details of the diagonalization of $\hat H_\mathrm{transfer}$ are given in Sec. III D in the supplementary material.

In the present case, the Gd(II) ion sits in a positive axial crystal-field created by the Gd(III) ion. The Gd(II) ion in its 4f$^8$ configuration is isoelectronic to a Tb(III) ion and based on purely electrostatic considerations, the lowest single-ion state is the singlet with $M_J=0$.\cite{long_2011a} The 16-fold degenerate direct product states involving the $M_J = 0$ crystal-field state are split into eight Kramers doublets by $\hat{H}_\text{transfer}$ (Table \ref{T:4f8_transfer_states}). The splitting is much weaker than in the case where the Gd(II) ion has a 4f$^7$5d$^1$ configuration as is expected from the much shorter spatial extent of the 4f orbitals as compared to the 5d orbitals. The largest splitting is obtained for $|M_{J0}|=1$ in \textbf{2} which is still only $25.0\,\mathrm{cm}^{-1}$.

This confirms that the interaction between states characterized by different value of $|M_{J0}|$ can be safely neglected when the splitting between the $|M_{J0}|$ states is large. In the $M_{J0}=0$ case, because of the axiality, $\hat{H}_\text{transfer}$ merely exchanges the local $M_{J0}$ and $M_0$ projections on the two sites, and the total projection $M_{J0}+M_0$ is conserved. For other values of $M_{J0}$ in the present system ($S_0=7/2$, $M_{J0}$ integer) the same projections are also conserved in the cases when $|M_{J0}|>3$. For values $0<|M_{J0}|\leq 3$, the $M_{J0}$ and $-M_{J0}$ components of the doublets mix and only the total angular momentum projection $M_J$ is conserved. In the $M_{J0}=0$ and $|M_{J0}|>3$ cases the splitting of the various single-ion states is linearly proportional to the transfer parameters and the spacing between the energy levels depends only on the transfer parameters and the angular momentum projections $M_{J0}$ and $M_0$ (see Table \ref{T:4f8_transfer_states} for $M_{J0} = 0$ and Table S8 in the supplementary material for $|M_{J0}|>3$). The transfer interaction is thus purely of Ising type. In the $0<|M_{J0}|\leq 3$ case the interaction is still of Ising type but the dependence on the transfer parameters becomes more complicated due to the mixing of the states characterized by the $M_{J0}$ and $-M_{J0}$ quantum numbers.

\begin{table}
  \caption{Splitting of the $M_{J0}=0$ state arising from the 4f$^8$ configuration in \textbf{1} and \textbf{2} due to 4f$\leftrightarrow$4f electron transfer}
  \label{T:4f8_transfer_states}
  \begin{ruledtabular}
    \begin{tabular}{crrr}
      $M_0$ & Exact expression & \textbf{1} / $\mathrm{cm}^{-1}$ & \textbf{2} / $\mathrm{cm}^{-1}$ \\
      \hline
      $\pm 1/2$ & $-\frac{25}{7392}\left(64t_0 + 27t_1\right)$ & $ -8.215$ & $-13.915$ \\
      $\pm 3/2$ & $- \frac{3}{2464}\left(125t_1 + 8t_2\right)$ & $ -3.549$ & $ -6.220$ \\
      $\pm 5/2$ & $  -\frac{1}{7392}\left(216t_2 + t_3\right)$ & $ -0.105$ & $ -0.185$ \\
      $\pm 7/2$ & $                        -\frac{1}{1056}t_3$ & $ -0.001$ & $  0.000$ \\
      $\pm 7/2$ & $                         \frac{1}{1056}t_3$ & $  0.001$ & $  0.000$ \\
      $\pm 5/2$ & $   \frac{1}{7392}\left(216t_2 + t_3\right)$ & $  0.105$ & $  0.185$ \\
      $\pm 3/2$ & $  \frac{3}{2464}\left(125t_1 + 8t_2\right)$ & $  3.549$ & $  6.220$ \\
      $\pm 1/2$ & $ \frac{25}{7392}\left(64t_0 + 27t_1\right)$ & $  8.215$ & $ 13.915$ \\
    \end{tabular}
  \end{ruledtabular}
\end{table}

\section{General discussion and conclusions}

The magnetic properties and the full spectrum of the lowest excited configuration of two endohedral metallo-fullerenes, [Gd$_2$@C$_{78}$]$^{-}$ (\textbf{1}) and [Gd$_2$@C$_{80}$]$^{-}$ (\textbf{2}), have been studied by theoretical methods. The two Gd ions have 4f$^7$ configurations with a single ``extra'' 5d electron delocalized between the ions. The spins of the unpaired electrons are coupled to an $S=15/2$ high-spin configuration. Depending on whether the 5d electron occupies a $\sigma$-symmetric orbital (as in the ground state) or a $\pi$- or $\delta$-symmetric orbitals, the configurations give rise to $\Sigma$, $\Pi$ or $\Delta$ terms, each of which splits very differently under exchange, spin-orbit and electron transfer interactions leading to a multitude of different energy level structures in the final spectrum.

In the ground $\Sigma$ manifold the transfer interaction is very strong ($t\ll J_H'$) bordering on a covalent one-electron Gd--Gd bond. Thus, electron delocalization is the dominant effect in the spectrum of the $\Sigma$ states and Hund's rule coupling splits the delocalized states characterized by a total spin $S$. This leads to an energy level spacing which follows the Land\'e interval rule and the system can be described by a Heisenberg-type Hamiltonian. The interaction between the 4f spins of the Gd ions and the 5d electron is ferromagnetic and stabilizes the $S=15/2$ ground spin state. Due to the lack of spatial degeneracy in the $\Sigma$ states, all first order angular momentum effects are quenched and the total spin $S$ remains a good quantum number. Weak mixing of higher-lying manifolds into the $\Sigma$ states leads to zero-field splitting of the $^{2S+1}\Sigma$ terms which can be qualitatively described as a second order effect.

The situation is quite different in the excited $\Pi$ and $\Delta$ states where the 5d electron is promoted to a $\pi$ or $\delta$ symmetric orbital and first order angular momentum is not quenched. The splitting of the single-ion states due to Hund's rule coupling is stronger than the splitting due to SOC and therefore the single-ion energy levels emerging from Hund and non-Hund spin configuration retain two distinct manifolds. SOC, however, strongly mixes states corresponding to the same value of the total spin projection of the Gd(II) ion and therefore the quantum number $K$ describing the total spin of the ion is no longer a good quantum number. The transfer interaction further mixes the single-ion states and only the total angular momentum projection remains a good quantum number. In both the $\Pi$ and $\Delta$ states the transfer interaction is stronger than the splitting due to SOC and the single-ion spectra become strongly mixed resulting into a highly complicated energy level structure. In the $\Delta$ states the splitting due to electron transfer is weaker than the splitting due to Hund's rule coupling. In the absence of SOC this would be the condition for the validity of the conventional double exchange model. The $\Pi$ states start to approach the covalent limit where the splitting due to transfer interaction is stronger than the splitting due to Hund's rule coupling. At small magnitudes of the transfer parameter $|t|$, when an analytical expression can be given for the energies of the transfer-split states, it is clear that the splitting has a linear dependence on both the total single-ion spin $K$ of the Gd(II) ion and its projection $M_K$. This introduces an Ising-like dependence into the splitting which is very different from the splitting of states in the conventional isotropic double exchange situation where the splitting is proportional only to the total spin $S$ of the coupled system and leads to an energy level spacing proportional to $2S+1$.\cite{kahn_1993}

The splitting of single-ion energy levels under the electron delocalization was also examined in the case of states arising when the ``extra'' electron occupies a 4f orbital combination instead of a 5d orbital combination. In this case the SOC is assumed much stronger than the crystal-field splitting and the Gd(II) single-ion states are characterized by a total angular momentum $J_0$ and its projection $M_{J0}$. Different states corresponding to the same value of $|M_{J0}|$ split under the crystal field and the transfer interaction then splits these states. Two distinct cases are observed depending on the magnitude $|M_{J0}|$ of the angular momentum projection. If $M_{J0}=0$ or $|M_{J0}|>3$ no mixing between degenerate single-ion states corresponding to values $M_{J0}$ and $-M_{J0}$ takes place and the splitting is linearly proportional to the transfer parameters $\{t_{m_l}\}$. If $0<|M_{J0}|\leq 3$ the $M_{J0}$ and $-M_{J0}$ states become mixed and the dependence of the splitting on $\{t_{m_l}\}$ becomes more complicated. In both cases the splitting is of Ising type and very weak due to the strongly shielded nature of the 4f orbitals involved.

The results presented here provide a solid rationalization for the ferromagnetic $S=15/2$ ground state of \textbf{1} and \textbf{2} (and by extension to that of Gd$_2$@C$_{79}$N) in terms of the microscopic interactions which take place between the two Gd ions. In addition, we have discussed the magnetism of the excited states arising from configurations where the ``extra'' electron is promoted to a $\pi$ or $\delta$ symmetric 5d orbital combination or to a 4f orbital combination. These results reveal the unique nature of anisotropic electron transfer interactions. The present work constitutes the first detailed study of spin-dependent delocalization in the presence of first-order orbital angular momentum. It is worth emphasizing here that in all cases considered here where first-order orbital momentum is involved, the splitting of the energy levels is very different from that associated with conventional double exchange and the splitting will always have a direct dependence on the \emph{projection} of the angular momenta involved. The derivations presented here can be extended upon to describe the magnetic properties in analogous systems where the orbital contribution to the magnetic moment of the 4f electrons is not quenched (such as [Dy$_2$@C$_{78}$]$^{-}$) as well as to exchange-coupled mixed-valence lanthanide complexes. Therefore the present work also contributes to the future design and understanding of molecules with novel magnetic functionality.

\section*{Supplementary material}
See supplementary material for the details on geometry optimizations, the extraction of exchange coupling constants and ZFS parameters from broken symmetry DFT and DFT/ROCIS calculations, complete derivations of effective Hamiltonians, complete listings of numerical values of the exchange spectrum arising form the ground configuration, the extracted CF and transfer parameters, the splitting of $|M_{J0}|$ states by $4f\leftrightarrow 4f$ transfer interaction, and the optimized Cartesian coordinates.

\begin{acknowledgements}
We would like thank Prof. Tatsuhisa Kato (Kyoto University) for sending us his unpublished EPR data and for useful discussions.
A.\,M. gratefully acknowledges computational resources provided by prof. H.\,M. Tuononen (University of Jyv\"askyl\"a) and CSC-IT Center for Science, Ltd., in Finland as well as funding provided by the Academy of Finland (project No. 282499).
N.\,I. is a JSPS Overseas Research Fellow.
\end{acknowledgements}

\bibliography{computational,references}

\end{document}